\documentclass[a4paper]{jpconf}
\usepackage{graphicx}
\usepackage{bm}
\begin{document}

\newcommand{\be}{\begin{equation}}
\newcommand{\ee}[1]{\label{#1}\end{equation}}
\newcommand{\bem}{\begin{eqnarray}}
\newcommand{\eem}[1]{\label{#1}\end{eqnarray}}
\newcommand{\eq}[1]{Eq.~(\ref{#1})}
\newcommand{\Eq}[1]{Equation~(\ref{#1})}
\newcommand{\vp}[2]{[\mathbf{#1} \times \mathbf{#2}]}

\title{Observation of spin superfluidity: YIG magnetic films and beyond}

\author{Edouard Sonin}

\address{Racah Institute of Physics, Hebrew University of
Jerusalem, Givat Ram, Jerusalem 91904, Israel}

\ead{sonin@cc.huji.ac.il}

\begin{abstract}
From topology of the order parameter of  the magnon condensate observed in yttrium-iron-garnet (YIG) magnetic films one must not expect energetic barriers making spin supercurrents metastable. But we show that some barriers of dynamical origin are possible  nevertheless until the gradient of the phase (angle of spin precession)  does not exceed the critical value  (analog of the Landau critical velocity in superfluids).  On the other hand, recently published claims of experimental detection of spin superfluidity in YIG films and antiferromagnets  are not justified, and spin superfluidity in  magnetically ordered solids has not yet been  experimentally confirmed.
\end{abstract}

\section{Introduction}

Spin superfluidity  was suggested in the 1970s \cite{ES-78b,ES-82} (see recent reviews \cite{Adv,Mac}).  Manifestation of spin superfluidity is a stable spin supercurrent proportional to the  gradient of  the phase $\varphi$ (spin rotation angle in a plane) and  not accompanied by dissipation, in contrast to a dissipative spin diffusion current proportional to the gradient of spin density. 

In general the spin current proportional to the  gradient of  the phase $\varphi$ is ubiquitous and exists in any spin wave, although in these cases variation of the phase $\varphi$ is small \cite{HalHoh}. Analogy with mass and charge persistent currents (supercurrents) arises when  at long (macroscopical) spatial intervals along streamlines the  phase variation is many times larger than $2\pi$. 
The supercurrent state is a helical spin structure, but in contrast to equilibrium  helical structures is metastable.

An elementary process of relaxation of the supercurrent is phase slip. In this process a vortex with $2\pi$ phase variation around it crosses current streamlines decreasing the total phase variation along streamlines by $2\pi$. Phase slips are suppressed by energetic barriers for vortex creation, which disappear when phase gradients reach critical values determined by the Landau criterion.
Emergence of superfluidity is conditioned by special topology of the order parameter space. This requires the easy-plane anisotropy for the spontaneous magnetization in ferromagnets or the antiferromagnetic vector in antiferromagnets \cite{Adv,Mac}.

Experimentally evidence of spin superfluidity was obtained in the $B$-phase of superfluid $^3$He \cite{Bun}, in which spin degrees of freedom of the order parameter make it analogous to antiferromagnets.  Recently Bozhko {\em et al.} \cite{spinY}  declared  detection of spin supercurrent at room temperature in a coherent magnon condensate created in yttrium-iron-garnet (YIG) magnetic films by strong parametric pumping \cite{Dem6}. In YIG the equilibrium order parameter in the spin space is not confined to some easy plane as required   for  spin superfluidity.  Nevertheless, Sun {\em et al.}  \cite{Pokr} argued  that spin superfluidity is still possible.
Indeed, one cannot rule out that metastability of supercurrent states is provided by barriers not connected with topology of the equilibrium order parameter, if the system is far from the equilibrium. In this work we confirm that it is possible indeed  and determine critical values of possible supercurrents at which metastability is lost (the analogue of the Landau criterion).  

In the end we overview experiments on experimental  detection of spin superfluidity in various materials.  We define the term {\em superfluidity} in its original meaning known from  the times of Kamerlingh Ohnes and Kapitza: transport of some physical quantity (mass, charge, or spin) on macroscopical distances without essential dissipation. This definitely requires a large number of full $2\pi$ rotations along current streamlines as pointed out above.
On the basis of this we conclude that  recent claims of observation spin superfluidity in YIG films \cite{spinY} and in antiferromagnets \cite{BunLvo} were not justified, and further experiments are needed for observation of spin superfluidity in magnetically ordered solids.

\section{Topology and  superfluid spin currents } \label{TopCur}

\begin{figure}
\begin{center}
\includegraphics[width=20pc]{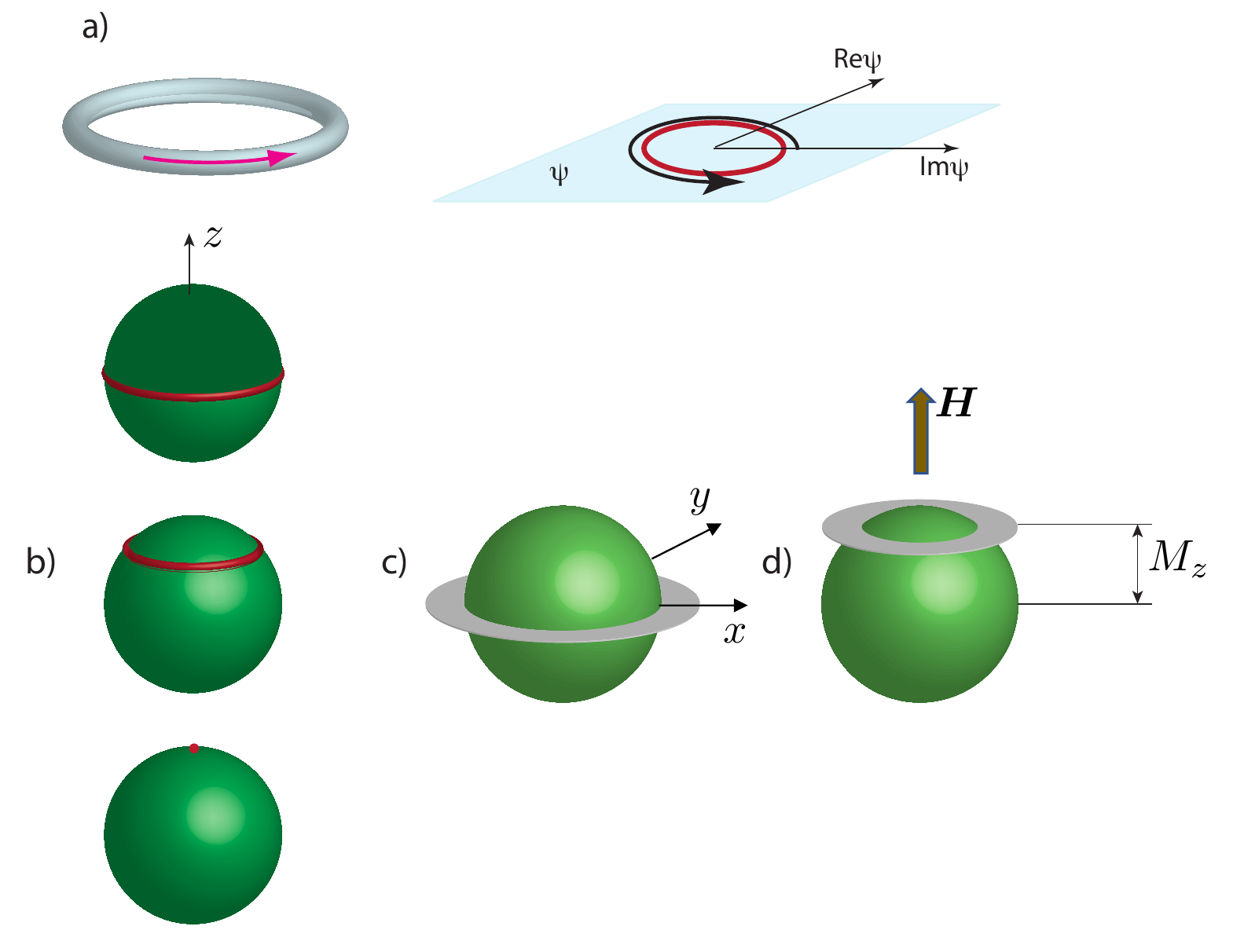}
\end{center}
\caption{\label{Fig02}Mapping of current states on the order parameter space.\newline
a) Mass currents in superfluids. The current state in torus maps on a circumference of radius $|\psi|$ on the complex plane $\psi$. \newline
b) Spin currents in an isotropic ferromagnet. The current state in torus maps on an equatorial circumference on the sphere of radius $M$ (top). Continuous shift of the path on the sphere (middle) reduces it to a point  (bottom), which is the ground state without currents. \newline
c) Spin currents in an easy-plane ferromagnet. Easy-plane anisotropy contracts the order parameter space to an equatorial circumference in the $xy$ plane topologically equivalent to the order parameter space in superfluids. \newline
d) Spin currents in an isotropic ferromagnet in a magnetic field $\bm H$ parallel to the axis $z$ with nonequilibrium magnetization $M_z$ supported by magnon pumping. Spin is confined in  the plane normal to $\bm H$. This  is an ``easy plane'' of dynamical origin.}
\end{figure}

In a superfluid the order parameter  is a complex wave function   $\psi = \psi_0 e^{i\varphi}$, where the modulus $\psi_0$ of the wave function is a positive constant determined by minimization of the energy. The
 phase $\varphi$ is a degeneracy parameter since the energy does not depend on  $\varphi$ because of gauge invariance. Any of degenerate ground states in a closed annular channel (torus) maps on some point at a circumference $|\psi|=\psi_0$ in the complex plane $\psi$, while a
 current state with  the phase change $2\pi n$ around the torus maps onto a circumference (Fig.~\ref{Fig02}a) winding around the  circumference $n$ times. It is evident that it is impossible to change $n$ keeping the path on the circumference $|\psi|=\psi_0$ all the time. In the language of topology states with different $n$ belong to different classes, and
 $n$ is a {\em topological charge}. Only a phase slip can change it when the path in the complex plane leaves  the circumference. This should cost energy, which is spent on creation of a vortex crossing the cross-section of the torus channel and changing $n$ to $n-1$.

If we consider transport of spin parallel to the axis $z$ the analog of the phase of the superfluid order parameter is the rotation  angle of the spin component in the plane $xy$, which we note also as $\varphi$.  In isotropic ferromagnets the order parameter space is a sphere of radius equal to the absolute value of the magnetization vector $\bm M$ (Fig.~\ref{Fig02}b). All points on this sphere correspond  to the same energy of the ground state.  Suppose we created the spin current state with monotonously varying  phase $\varphi$ in a torus. This state maps on the equatorial circumference on the order parameter sphere. Topology allows to continuously shift the circumference and to reduce it to the point of the northern pole.  During this process shown in Fig.~\ref{Fig02}b  the path remains on the sphere all the time and therefore no energetic barrier resists to the transformation. Thus metastability of the current state is not expected.

In a ferromagnet with easy-plane anisotropy the order parameter space contracts from the sphere to an equatorial circumference in the $xy$ plane. This makes the order parameter space topologically equivalent to that in superfluids (Fig.~\ref{Fig02}c).  Now transformation of the equatorial circumference to the point  shown in Fig.~\ref{Fig02}b costs anisotropy energy. This allows to expect metastable spin currents (supercurrents). They relax to the ground state via phase slips, in which magnetic vortices cross spin current streamlines. States with vortices maps on  a hemisphere of radius $M$ either above or below the equator.

Up to now we considered states close to the equilibrium (ground) state. In a ferromagnet  in a magnetic field the equilibrium magnetization is parallel to the field. However,  by pumping magnons into the sample it is possible to tilt the magnetization with respect to the magnetic field. This creates a nonstationary state, in which the magnetization precesses around the magnetic field. Although the state is far from the true equilibrium, but it, nevertheless, is a state of minimal energy at fixed magnetization $M_z$. Because of inevitable spin relaxation  the state of uniform precession requires permanent pumping of spin and energy. However, if  processes violating the spin conservation law are weak, one can ignore them and treat the state as a quasi-equilibrium state. The state of uniform precession maps on a circumference parallel to the $xy$ plane, but in contrast to the easy-plane ferromagnet (Fig.~\ref{Fig02}c) the plane confining the precessing magnetization is much above the equator and not far the northern pole (Fig.~\ref{Fig02}d).  One can consider also a current  state, in which the phase (the rotation angle in the $xy$ plane) varies not only in time but also in space with a constant gradient. The current state will be metastable due to the same reason as in an easy-plane ferromagnet:  in order to relax via phase slips the magnetization should go away from the circumference on which the state of uniform precession maps, and this increases the energy. Then the plane, in which the magnetization precesses, can be considered as an effective ``easy plane'' originating not from the equilibrium order parameter topology but created dynamically. The concept of dynamical easy plane is applicable for a magnon condensate   in YIG films with some modifications. They take into account that the spin conservation law is not exact due to magnetostatic energy and the precession is not uniform  since  spin waves in YIG films have the energy minima at non-zero wave vectors. 

In our discussion we assumed that phase gradients were small and ignored the gradient-dependent (kinetic) energy. At growing gradient  we reach the critical gradient at which barriers making the supercurrent stable vanish. The critical gradient must be determined from the. criterion analogous to  the famous Landau criterion \cite{Adv,Son17}. 

\section{ Spin waves in YIG films} \label{LLT}

The coherent state of magnons is nothing else but a classical spin wave, and one can use the classical equations of   the Landau--Lifshitz--Gilbert (LLG) theory.
YIG  is a  ferrimagnet with complicated magnetic structure consisting of numerous sublattices. However at  slow degrees of freedom relevant for our analysis one can treat it simply as an isotropic ferromagnet \cite{Melk} with the spontaneous magnetization $\bm M$ described by the hamiltonian
\bem
{\cal H }=\int\left[-\bm H\cdot \bm M +D{\nabla_i  \bm M  \cdot \nabla_i  \bm M\over 2} \right]d\bm r
+ \int {\bm \nabla \cdot \bm M(\bm r) \bm \nabla \cdot \bm M(\bm r_1)\over 2|\bm r-\bm r_1| } d\bm r\,d\bm r_1 .
  \eem{}
Here the first term is the Zeeman energy in the magnetic field  $\bm H$, the second term  $\propto D$ is the inhomogeneous exchange energy, and  the last one is the magnetostatic (dipolar) energy.

In a weak spin wave  propagating in the plane $xz$ in  a magnetic field $\bm H$ parallel to the axis $z$
$M_z \approx M-{M_\perp^2/2M}$ and $\bm \nabla \cdot \bm M \approx \nabla_x M_x$,
where $M_\perp=\sqrt{M_x^2+M_y^2}$.  In  the LLG theory the absolute value of the magnetization vector $\bm M$ does not vary in space and time, and the linearized LLG equations are reduced to two equations  for only two independent magnetization components:
 \bem
\dot M_x=-\gamma M{\delta {\cal H} \over \delta M_y}=-\gamma H M_y + \gamma DM (\nabla_x^2 M_y+\nabla_z^2 M_y),
\nonumber \\
\dot M_y=\gamma M_z {\delta {\cal H} \over \delta M_x} =\gamma H M_x - \gamma DM (\nabla_x^2 M_x+\nabla_z^2 M_x)
-  \gamma M\nabla_x \left(\int {\nabla_x M_x(\bm r_1) \over |\bm r-\bm r_1| }d\bm r_1 \right),
     \eem{em}
where  $\gamma$ is the gyromagnetic ratio and $\delta {\cal H} / \delta M_x$ and $\delta {\cal H} / \delta M_y$ are functional derivatives. For the plane wave with the frequency $\omega$ and the wave vector $\bm k(k_x,0,k_z)$ Eqs.~(\ref{em}) yield the spectrum
\be
\omega(k)= \gamma \sqrt{(H + DM k^2)\left(H + DM k^2 +{4\pi M k_x^2\over k^2}\right)}.
   \ee{disp}

\begin{figure}
\includegraphics[width=15pc]{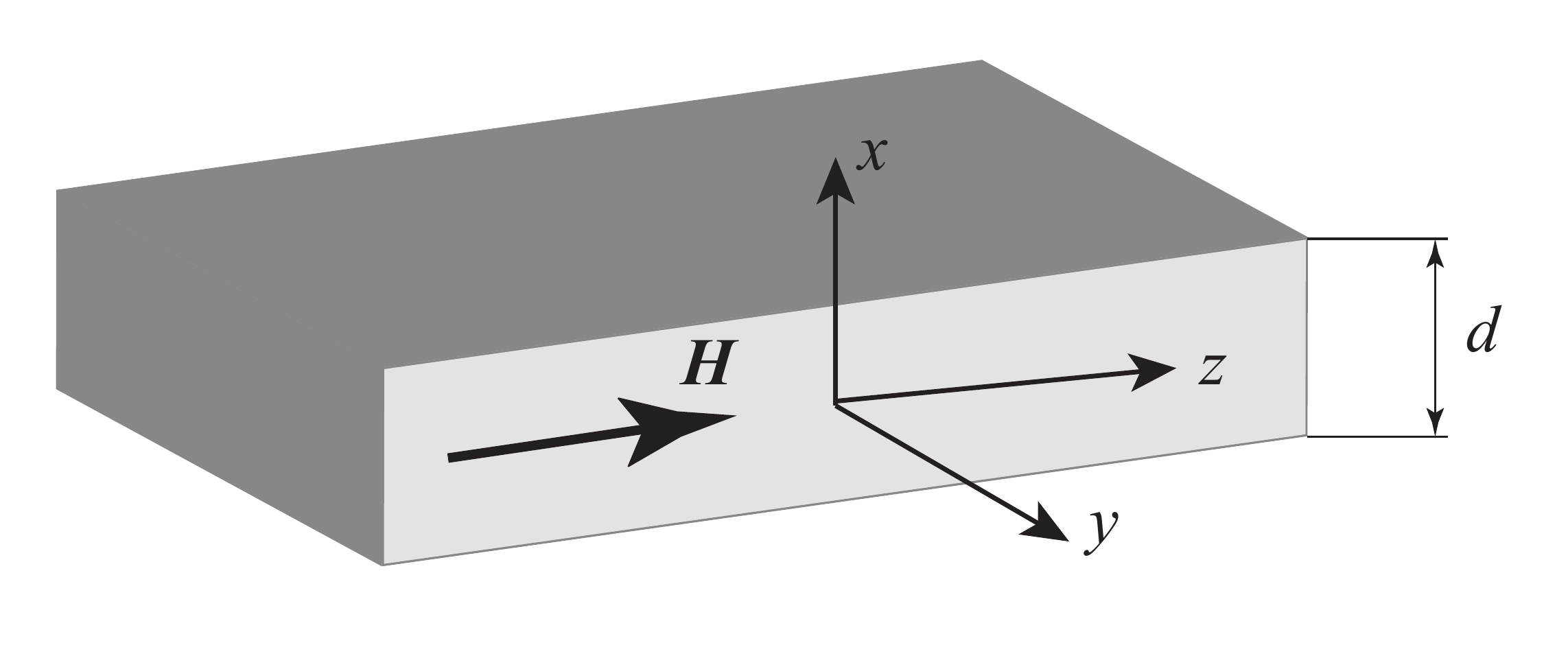}\hspace{2pc}
\begin{minipage}[b]{14pc}\caption{\label{Fig1}The YIG film of thickness $d$ in a magnetic field $\bm H$ parallel to the axis $z$.}
\end{minipage}
\end{figure}

A spin wave propagating in the film  of thickness $d$ parallel to the plane $yz$ (Fig.~\ref{Fig1})  must satisfy the boundary conditions at two film surfaces $x=\pm d/2$. In the situation when both the exchange and the dipole energies are of importance, the boundary problem is not trivial and was discussed in details in Ref.~\cite{Son17}.  According to this  analysis, the spectrum of the spin wave propagating along the magnetic field in the film is
\be
\omega(k_z) \approx  \gamma \left(H + DM k_z^2 +{2\pi^3 M \over k_z^2d^2}\right).
   \ee{disp1}
This follows from the spectrum \eq{disp}  at $k_x= \pi /d$ and corresponds  to
a plane wave with weakly elliptical polarization propagating along the magnetic field (the axis $z$).
The energy $E =  \omega (M-\langle M_z\rangle)/ \gamma$
and the frequency $\omega(k_z)$ given by \eq{disp1} have two degenerate minima \cite{Melk}  at finite $k_z =\pm k_0$ where magnons can condense  (Fig.~\ref{Fig2}). Here
\be
k_0=\left({2\pi^3\over Dd^2}\right)^{1/4}=\left({2\pi^2\over l_d^2 d^2}\right)^{1/4},
     \ee{}
where $l_d=\sqrt{D/\pi}$ is a small scale determined by the exchange energy.

\section{Spin supercurrent in the YIG magnon condensate}

In the linear theory  the distribution of magnons between two condensates is arbitrary and does not affect the total energy at fixed magnetization $\langle M_z\rangle$.  But  non-linear corrections lift this degeneracy.  The most important nonlinear term arises from the magnetostatic energy, which is maximal  
for the standing wave (two energy minima are equally populated by magnons): 
\bem
M_x =2\sqrt{2M(M-\langle M_z\rangle)}\cos {\pi x\over d} \cos k_zz\cos \omega t, 
\nonumber \\
 M_y = 2\left(1+{2\pi^3 M\over Hk_z^2d^2}\right)\sqrt{2M(M-\langle M_z\rangle)}\cos {\pi x\over d}\cos k_zz \sin\omega t,
    \eem{sw}
 and is equal to
\be
E_{ms}  =\int {\nabla _z M_z(\bm r)  \nabla_z M_z(\bm r_1))\over 2|\bm r-\bm r_1| } d\bm r\,d\bm r_1={3 \pi (M-\langle M_z\rangle)^2 \over 2}.
    \ee{mse}

\begin{figure}[b]
\includegraphics[width=15pc]{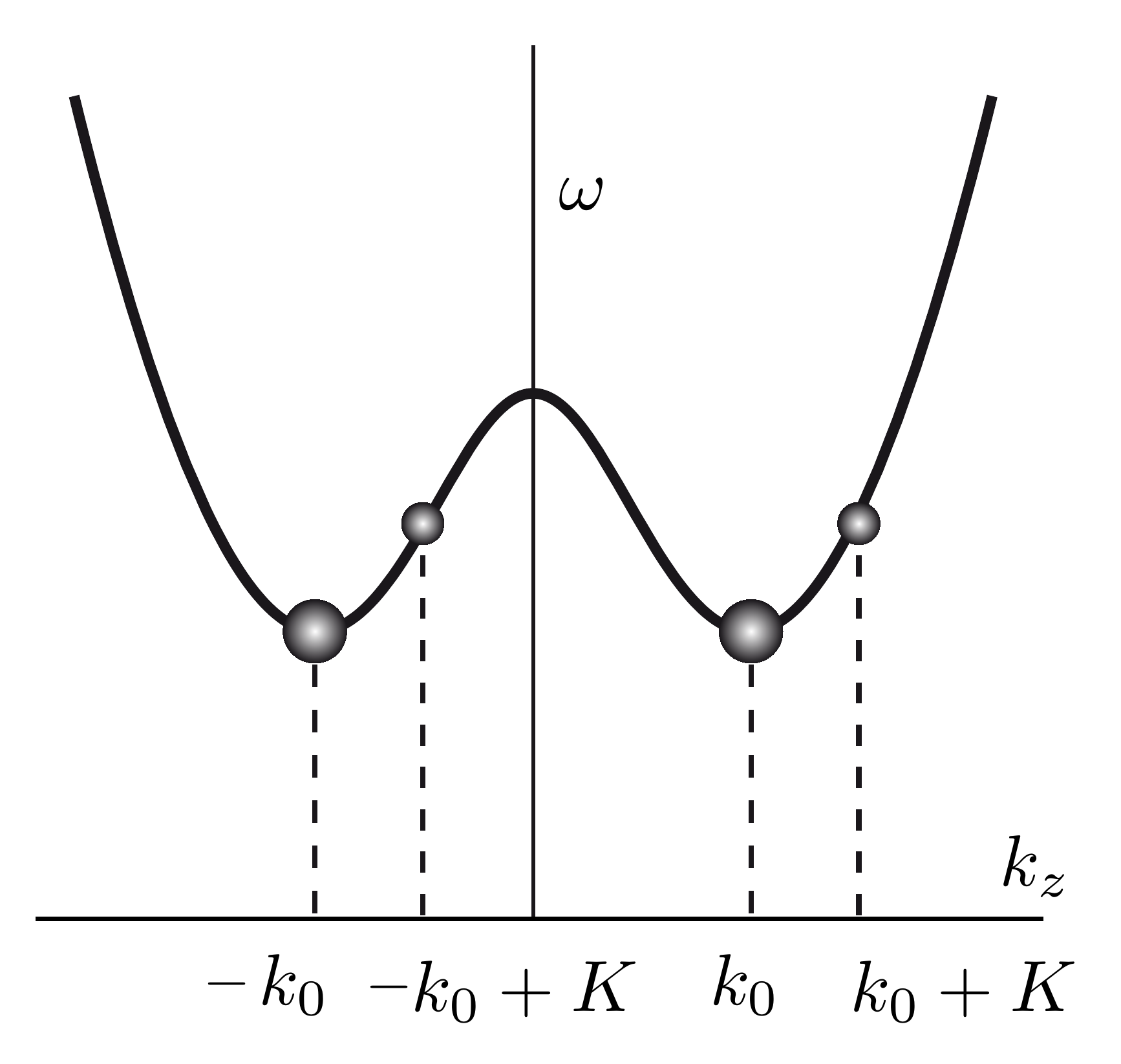}
\begin{minipage}[b]{14pc}\caption[]{\label{Fig2}The spin-wave spectrum in a YIG film. In the ground state the magnon condensate occupies two minima in the $k$ space with $k_z=\pm k_0$ (large circles). In the current state two parts of the condensate are shifted to $k = \pm k_0 +K$ (small circles). }
\end{minipage}
\end{figure}

The spin current appears if the wave numbers $k_z$  of two condensates differ from $\pm k_0$ (Fig.~\ref{Fig2}). The current is proportional to the phase gradient  $\nabla_z\varphi=K=k_z-k_0 \ll k_0$. Keeping the magnetization $\langle M_z \rangle$  fixed as before and taking into account the nonlinear magnetostatic term (\ref{mse})  the  energy in the spin-current state apart from some constant terms  is
 \be
 \Delta E = {d^2\omega (k_z)\over dk_z^2} {M-\langle M_z\rangle\over \gamma}{(\nabla_z\varphi)^2\over 2}+{3 \pi (M-\langle M_z\rangle)^2 \over 2}.
        \ee{hMz}

Stability of the spin-current state can be checked following the principal idea of the Landau criterion of superfluidity \cite{Adv,Son17}.
Let us consider slowly varying in space weak perturbations $m_z=M_z-\langle M_z \rangle $ and $\nabla_z\varphi'=\nabla_z \varphi-K $. For stability of the supercurrent the quadratic form obtained from expansion of the energy (\ref{hMz})   in perturbations   $m_z$ and $\nabla_z \varphi'$ must be always positive. This takes place as far as $\nabla_z\varphi=K$ is less than the critical value
\be
(\nabla_z\varphi)_{cr} =\sqrt{3\pi \gamma (M-\langle M_z \rangle)\over {d^2\omega (k_0)\over dk_z^2} }=  \sqrt{{ 3(M-\langle M_z \rangle)\over M}}{ k_0^2 d\over 4\pi }.
     \ee{}
Note that applying our course of derivation to superfluid hydrodynamics one obtains exactly the Landau critical velocity equal to the sound velocity (see Sec.~2.1 in Ref.~\cite{Adv}). One can find numerical estimation of the critical gradient and comparison with other theories in Ref.~\cite{Son17}.

\section{Experimental detection of spin superfluidity} \label{DC}
 
A smoking gun of spin superfluidity in the $B$-phase of superfluid $^3$He was  an experiment with a spin current through a long channel connecting two cells filled by the fluid with coherently precessing spins \cite{BRJ}. A small difference of the precession frequencies in two cells  leads to a linear growth of
difference of the precession phases in the cells and a phase gradient in the channel. When the gradient reaches the critical value phase slips must occur. Phase slips were detected experimentally at the total phase difference exceeding $2\pi$. 
A sharp $2\pi$  phase slip  is reliable evidence of non-trivial spin supercurrents at phase gradients restricted by finite critical values. 

Some time ago Bunkov {\em et al.} \cite{BunLvo} declared that their observation of coherent spin precession in antiferromagnets demonstrated  ``high-$T_c$ spin superfluidity'' since random inhomogeneity of their samples  inevitably  generates local spin currents. The very idea that random spin currents  
fluctuating at the disorder scale are the same as persistent currents, which are constant at macroscopic scales, is bizarre at least. Moreover,  no shred of evidence, direct or indirect, of presence of any spin current, random or non-random, was presented. The spin current, or any quantity connected with it,  is totally absent in any experimental data or formula. The only ``evidence''  was the statement that spin currents {\em must} be present in their experiment. Apparently  Bunkov {\em et al.} do not see any difference between a theoretical prediction ({\em must} be present) and experimental confirmation of the prediction.

Recently  Bozhko {\em et al.}  \cite{spinY} declared detection of spin superfluidity at high temperatures in a decaying magnon condensate in a YIG film.   In their experiment the phase gradient emerged from  spin precession difference produced by a temperature gradient. But the estimate made in Ref.~\cite{Son17} showed that the total phase difference across the magnon cloud in the experiment  did not exceed  about 1/3 of the full $2\pi$ rotation. Spin current relaxation via discrete $2\pi$ phase slips is impossible in this  case independently from  phase gradient. Therefore Bozhko {\em et al.} dealt with trivially stable spin currents present in any spin wave. Their existence does not require new confirmations  after a half-a-century experiments  with spin waves at {\em all} temperatures.

In summary,  spin superfluidity, which was revealed  in superfluid $^3$He-$B$, 
  still waits its experimental confirmation in magnetically ordered solids.
 
\section*{References}
\providecommand{\newblock}{}

\end{document}